\documentclass[twocolumn]{aastex631}

\usepackage{enumitem}

\usepackage{array}
\usepackage{graphicx}
\usepackage{tabularx}
\usepackage{multirow}
\usepackage{amsmath}

\begin{document}

\title{Disparities in Magnetic Cloud Observations Between Two Spacecraft Having Small Radial and Angular Separations Near 1 AU}

\author[0009-0007-4956-5108]{Anjali Agarwal}
\affiliation{Indian Institute of Astrophysics, II Block, Koramangala, Bengaluru 560034, India}  
\affiliation{Pondicherry University, R.V. Nagar, Kalapet 605014, Puducherry, India}

\author[0000-0003-2740-2280]{Wageesh Mishra}
\affiliation{Indian Institute of Astrophysics, II Block, Koramangala, Bengaluru 560034, India}; 

\correspondingauthor{Anjali Agarwal and Wageesh Mishra}
\email{anjaliagarwal1024@gmail.com; m.wageesh30@gmail.com}

\begin{abstract}

Studies for inferring the global characteristics of coronal mass ejections (CMEs) from its multipoint local in situ observations have been undertaken earlier, but there are limited studies utilizing measurements from multiple spacecraft with sufficiently small radial and angular separations. In the present study, we investigate a magnetic cloud (MC) region of a CME observed in situ during 2023 September 24-26, at \textit{STEREO-A} and \textit{Wind} spacecraft near 1 AU, which had radial and angular separations of 0.03 AU and 3.4$^\circ$, respectively. We examine the disparities in the estimates of the arrival times of CME substructures, the MC axis, and its orientation between the two spacecraft. We also propose an approach for identifying the MC axis's arrival and have compared it with the arrival of the size/time center to understand the non-isotropic compression of the MC along its angular extent. Using minimum variance analysis (MVA), we note that the orientation of the MC is slightly out-of-ecliptic at \textit{Wind} but not at \textit{STEREO-A}. We also compare the magnetic field parameters over the start to end of the MC at both spacecraft and note a significant non-coherency in the MC towards its trailing portion. Our analysis confirms that MC has a stronger rear side compression at \textit{STEREO-A} than at \textit{Wind}, with its trailing edge arriving later at \textit{Wind}. Our study highlights substantial differences in CME characteristics even at mesoscales across the angular extent, and therefore, one needs to analyze several such cases to better understand the flux rope structure.

\end{abstract}


\section{Introduction} \label{sec:int}

Coronal Mass Ejections (CMEs) are energetic expulsions of magnetized plasma bubbles from the Sun's corona into the heliosphere and are major sources of space weather effects \citep{Schwenn2006,Pulkkinen2007,Webb2012,Schrijver2015}. CMEs are often observed using remote and in situ observations, and these two sets of observations have inherent advantages and limitations. There have been several attempts made to understand the global morphology, structures, orientation, and evolution of 3D kinematics and thermodynamics of CMEs in the heliosphere using remote and in situ observations combined with modeling \citep{Sheeley1999,Xie2004,Thernisien2006,Davies2013,Mishra2015,Mishra2018,Mishra2023a,Khuntia2023}. The in situ observations of a CME have been used to clearly identify its substructures, such as shock, sheath, and flux rope that is identified as magnetic cloud (MC), as their plasma properties are inherently different from each other \citep{Zurbuchen2006,Mishra2023,Temmer2024}. However, the possibility of identifying different substructures depends on the 1D trajectory of the in situ spacecraft through the global CME structure \citep{Burlaga1981,Crooker1996,Bothmer1998,Nieves-Chinchilla2013}. For example, only 30-40\% of CMEs exhibit flux rope (i.e., magnetic cloud) structure in the in situ observations \citep{Gosling1990,Song2020,Mishra2021a}, although it does not mean that flux ropes are absent in CMEs. It is considered that the signatures of missing flux rope in CMEs are due to the absence of a favorable trajectory of the spacecraft through the CME \citep{Gopalswamy2006,Zhang2013}. Therefore, observations from multiple in situ spacecraft can be inconsistent in terms of detecting some substructures, and further, the in situ observations of the magnetic field and plasma parameters of the same substructures from multiple viewpoints can provide their 3D morphology and structures \citep{Kilpua2011,Lugaz2018,Mishra2021}.

From the space weather perspective, the magnetic field and plasma parameters of CME's different substructures can play an important role in deciding their consequences on the Earth \citep{Gonzalez1999,Wang2003,Srivastava2004,Wood2017}. Furthermore, these parameters are expected to be different at different radial distances from the Sun as CMEs evolve in the ambient solar wind during their interplanetary (IP) journey from the Sun to 1 AU. Further, some studies have suggested that CME properties measured at longitudinally separated in situ spacecraft can differ, even if the multiple spacecraft are on the same plane and at the same radial distance from the Sun. This could be due to certain geometrical structures of CME and its flux rope, non-coherent magnetic field structures inside CMEs, inhomogeneity in the background solar wind, and non-isotropic expansion of CMEs \citep{Gopalswamy2006,Owens2017,Cremades2020,Desai2020,Kay2021,Mishra2021}. \citet{Lugaz2022} found that the western leg of a CME is moving slower than the eastern leg of the CME, which results in the formation of shock at the western leg, and suggested that the local parameters of CMEs as observed by in situ spacecraft are also influenced by the surrounding ambient medium. Additionally, \citet{Al-haddad2025} has emphasized the complexity of CMEs substructures, their deformation in the IP medium, and the necessity of multipoint in situ observations for a better understanding and representation of CMEs. Therefore, multipoint in situ spacecraft observations are required to understand the CME parameters globally and investigate the physics of CME evolution.

In the era of Helios, IMP, Pioneer, and Voyager spacecraft, studies utilizing multipoint in situ observations have suggested that CMEs can have larger expansion closer to the Sun and their flux ropes (MCs) can have highly distended cross-sections (longitudinal dimension) in the IP medium with respect to CMEs radial size \citep{Burlaga1981,Crooker1996,Bothmer1998}. However, such studies utilized observations from in situ spacecraft that were largely separated radially as well as longitudinally from each other. Therefore, such studies could not discern the effects of CME's temporal evolution with its inherent inhomogeneity on the measured plasma parameters at different spacecraft. This is because CME plasma parameters observed by radially aligned in situ spacecraft can differ due to their temporal evolution, while they can also differ at the longitudinally separated in situ spacecraft due to the inherent inhomogeneity in the CME and its non-coherent evolution. Moreover, these studies could be compromised in the absence of imaging observations as it would be difficult to ascertain if the same CME substructures were sampled at different locations of multiple in situ spacecraft.

The launch of twin \textit{Solar TErrestrial RElations Observatory} (\textit{STEREO}) \citep{Kaiser2008,Howard2008} with planetary missions \textit{MESSENGER} and \textit{Venus Express} provided the opportunity to analyze multipoint in situ measurements of CMEs at radially and longitudinally separated locations. Using radially aligned multipoint in situ spacecraft (with small separation in longitude), there are several attempts made to study the radial/temporal evolution of the CME and to determine the power law of the radial width of the CME and strength of the axial magnetic field with heliocentric distance \citep{Liu2005,Leitner2007,Gulisano2010,Good2015,Good2018,Lugaz2020,Salman2020,EmmaDavies2021}. Additionally, utilizing the temporal evolution and kinematics of the CME in the IP medium, there are models to examine the thermodynamic properties of the CME \citep{Mishra2018,Mishra2020,Khuntia2023}. Moreover, utilizing the new era spacecraft \textit{Parker Solar Probe} (\textit{PSP}), \textit{Solar Orbiter} (\textit{SolO}), and \textit{BepiColombo}, which are often separated by large radial distances and small/large longitudes, there are studies exploring the radial/temporal evolution of CMEs and its substructures in the background solar wind medium \citep{EmmaDavies2021a,Kilpua2021,Winslow2021,Mostl2022,Regnault2023}. The interaction of CMEs with other large-scale solar wind structures (e.g., heliospheric current sheet, preceding CMEs, and stream interaction regions) can also lead to changes in their properties at different spatial scales \citep{Gopalswamy2009a,Winslow2016,Mishra2017,Kilpua2019,Mishra2021,Lugaz2022}.

It is obvious that there are only a handful of studies exploring CME characteristics measured by multiple spacecraft located at nearly the same distance from the Sun but with a small longitudinal separation \citep{Kilpua2011,Lugaz2018}. Thus, there remains a gap in exploring the CME properties at small spatial/mesoscale, which is possible if multiple spacecraft sampling the CME have mutual radial separations of 0.005-0.05 AU and longitudinal separations of 1$^\circ$-12$^\circ$ \citep{Lugaz2018}. Such observations at the mesoscale can emphasize the inherent inhomogeneity of CMEs or their different substructures. It is possible that the temporal evolution of each mesoscale substructure of CMEs is different, and they go through different physical processes during their interplanetary (IP) journey. Therefore, it is necessary to conduct a comprehensive study on CME substructures observed by two or more in situ spacecraft separated by only a small spatial/longitudinal extent, which can reveal the physics of CMEs at the mesoscale.

In this paper, we focus on the difference in the magnetic field and plasma parameters of a selected CME flux rope at mesoscales measured by \textit{STEREO-A} \citep{Luhmann2008, Galvin2008} and \textit{Wind} \citep{Ogilvie1997} as these two spacecraft provide a rare and favorable conjunction for such a study. The selected CME is found to arrive near 1 AU on 2023 September 24-26, during which the radial and angular separation between both spacecraft are ~0.03 AU ($\sim$6.4 $R_{\odot}$) and $\sim$3.4$^\circ$, corresponding to the arc length of around 0.06 AU, taking them to be at 0.99 AU, respectively. We also focus on a novel approach to determine the flux rope's axis (termed as the axis center), utilizing the measured magnetic field parameters of the flux rope (MC) at both spacecraft. We attempt to find the differences in the time, size, and axis centers of the MC at both spacecraft. We also examine the differences in the orientation of the CME flux rope axis estimated by utilizing the minimum variance analysis (MVA) technique on the observations of both spacecraft. The multipoint in situ observations of the CME, our approach to determining the flux rope's axis, MVA analysis, and mesoscale differences in the magnetic field parameters are described in Section~\ref{sec:insitu}, \ref{sec:centers}, \ref{sec:MVA}, and \ref{sec:mag_comp}, respectively. Section~\ref{sec:result} summarizes our results and discusses the factors that can lead to some uncertainties in our findings.

\section{OBSERVATIONS AND ANALYSIS METHODOLOGY}{\label{sec:observation}}

We focus on investigating the mesoscale differences in CME magnetic field and plasma parameters measured by multipoint in situ spacecraft separated by a small angular extent of around 3.4$^\circ$. The in situ observations are from \textit{STEREO-A} (\textit{STA}) and \textit{Wind} spacecraft during 2023 September 24-26. In the following, first, the in situ observations are analyzed to scrutinize the differences in the arrival time of CME substructures, such as shock, sheath, leading edge (LE), and trailing edge (TE), at both spacecraft. The differences in the duration of CME substructures, such as sheath and MC on both spacecraft, are also estimated. Further, we focus on introducing a new approach to determining the flux rope's/MC's axis, termed the axis center using in situ magnetic field measurements. The estimated axis center from the new approach, when compared with size and time center, can shed insights into the compression of the MC. The orientation of the MC axis on both in situ spacecraft is estimated using the minimum variance analysis (MVA) technique \citep{Sonnerup1967} to analyze the differences in the orientation along a small angular extent of the MC. We also calculate the root mean square error and mean absolute error for the differences between measurements at both spacecraft along the RTN axes and the variance axes determined by MVA. Additionally, we compute the Spearman (Pearson) correlation coefficient between measurements at both spacecraft along these axes.

\subsection{In Situ Observations of the CME Substructures at STEREO-A and Wind}\label{sec:insitu}

\begin{figure*}
    \centering
    \includegraphics[scale= 0.385,trim={0cm 0cm 0cm 0cm},clip]{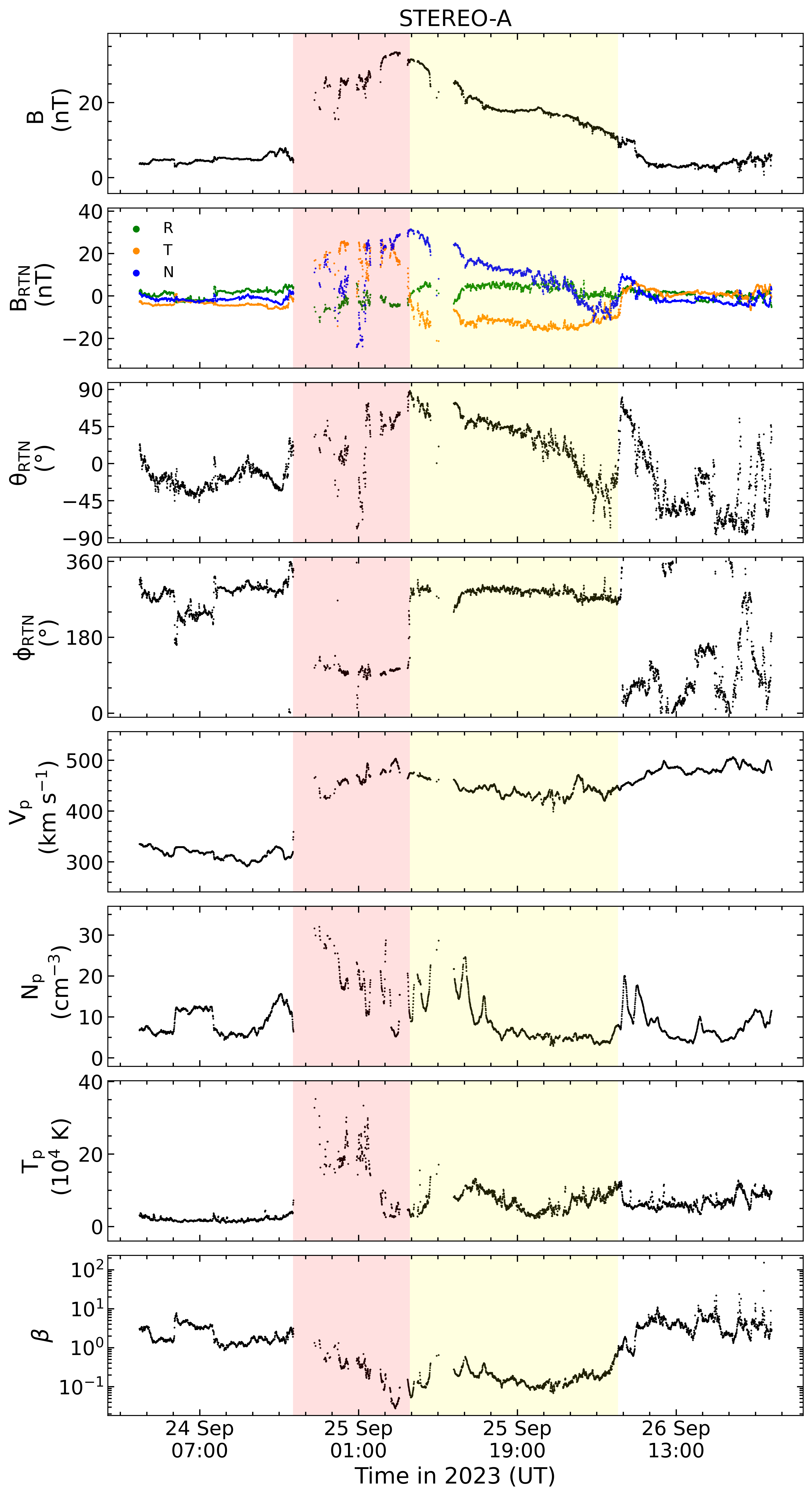}
    %\hspace*{1.5cm}
   \includegraphics[scale=0.385,trim={0cm 0cm 0cm 0cm},clip]{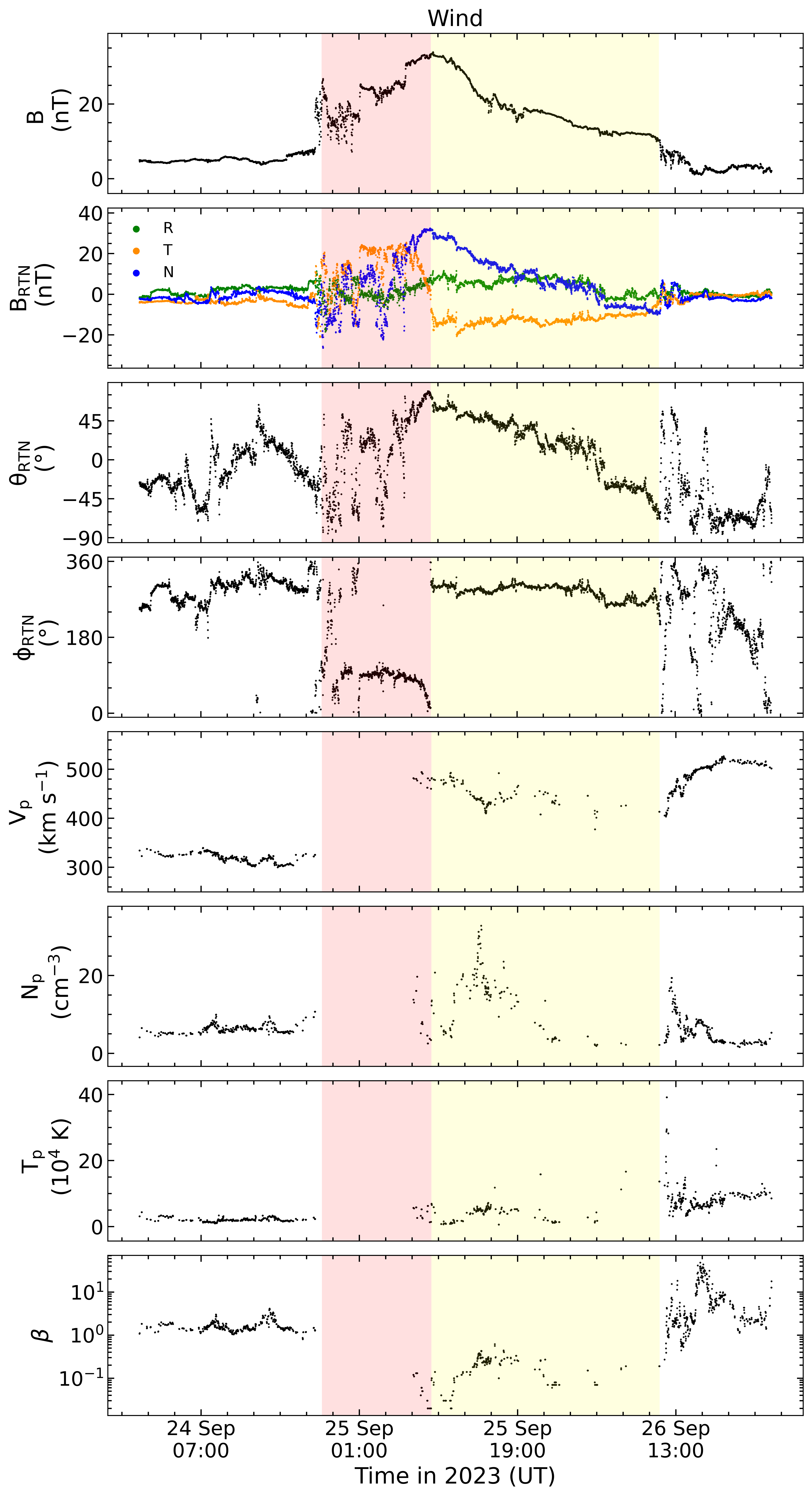}
    \caption{ The top to bottom panels show the variation of the total magnetic field, magnetic field vector in RTN coordinate system, latitude and longitude of the total magnetic field vector, speed, density, temperature, and plasma beta, respectively, from \textit{STA} and \textit{Wind} in the left and right column, respectively. Transparent fill areas with red and yellow represent the sheath and magnetic cloud duration during the passage of the CME on the spacecraft.}
    \label{fig:insitu_par}
\end{figure*}

The longitudinal and latitudinal separation between \textit{STA} and \textit{Wind} on 2023 September 25 are 3.4$^\circ$ and 0.1$^\circ$, respectively. We scrutinize the magnetic field and plasma parameters of selected CME for \textit{STA} and \textit{Wind} as depicted in the left and right column of Figure~\ref{fig:insitu_par}, respectively. In this figure, the top-to-bottom panels show the total magnetic field, radial, tangential, and normal components of the magnetic field vector (in the green, orange, and blue), latitude of the total magnetic field vector ($\theta$), longitude of the total magnetic field vector ($\phi$), speed, density, temperature, and plasma beta, respectively. We estimate $\theta$ using the total magnetic field and normal component of the magnetic field (${B_N}$) as follows:

$${\theta = \sin^{-1}(\frac{B_N}{B})}$$

We estimate $\phi$ in the RTN coordinate system using magnetic field components ${B_R}$ and ${B_T}$ as following:

if ${B_R>0}$ and ${B_T>0}$:
\begin{displaymath}
{\phi = \tan^{-1}\left(\frac{B_T}{B_R}\right)};~({0<\phi<\pi/2}) 
\end{displaymath}

if ${B_R<0}$ and ${B_T>0}$:
\begin{displaymath}
{\phi = \tan^{-1}\left(\frac{B_T}{B_R}\right) + 180^\circ};~({\pi/2<\phi<\pi})    
\end{displaymath}

if ${B_R<0}$ and ${B_T<0}$:   
\begin{displaymath}
{\phi = \tan^{-1}\left(\frac{B_T}{B_R}\right) + 180^\circ};~({\pi<\phi<3\pi/2})  
\end{displaymath}

if ${B_R>0}$ and ${B_T<0}$:
\begin{displaymath}
{\phi = \tan^{-1}\left(\frac{B_T}{B_R}\right) + 360^\circ};~({3\pi/2<\phi<2\pi})   
\end{displaymath}

The boundaries of CME substructures are identified based on their distinct signatures in the in situ observations near 1 AU \citep{Zurbuchen2006,Richardson2010}. However, for identifying the in situ boundaries at \textit{Wind}, we primarily rely on magnetic field measurements due to the data gap in plasma measurements. The start and end boundary of the MC at \textit{Wind} are 2023 September 25 at 09:12 UT and 2023 September 26 at 11:10 UT, respectively, which are slightly different than the start (2023 September 25 at 09:00 UT) and end (2023 September 25 at 11:00 UT) boundary of the MC in Richardson and Cane ICME catalog (\url{https://izw1.caltech.edu/ACE/ASC/DATA/level3/icmetable2.htm}). The slight difference in the selected boundaries compared to the catalog boundaries is because we strictly consider the rotation of magnetic field vectors (represented by $\theta$ and $\phi$) for marking the start and end boundary of the MC.

In Figure~\ref{fig:insitu_par}, the transparent fill areas with red and yellow depict the sheath and MC region. Also, the rotation of $\theta$ and $\phi$ within the MC boundary outline the orientation of the MC as North-East-South (NES) at both spacecraft \citep{Bothmer1998}. This orientation suggests that this MC is a low-inclined flux rope whose axis is in the east direction \citep{Bothmer1998, Palmerio2018}. The $\theta$ profile of the MC at both spacecraft shows that the rotation of $\theta$ is slow initially but suddenly decreases towards the end of the MC, and then remains nearly constant for some time. This type of variation in $\theta$ is more pronounced at \textit{Wind}. Moreover, this can happen due to the compression of the MC, as reflected by the speed measurements in the MC region of the fifth panel of Figure~\ref{fig:insitu_par}. In Section~\ref{sec:centers}, we examine the compression of the MC on both spacecraft.

The inspection of in situ measurements notes the arrival time of the shock, LE, and TE of the MC measured at \textit{STA} and \textit{Wind}, and the difference between both measurements ($\Delta t = t_{Wind} - t_{STA}$), which are listed in the second, third, and fourth column, respectively, in first panel of the Table~\ref{tab:tab_1}. We note that the arrival of CME substructures such as shock, LE, and TE at \textit{Wind} is later than their arrival at \textit{STA}. Such a late arrival at \textit{Wind} could be due to its larger distance from the Sun than that of \textit{STA} as both spacecraft are radially separated by 0.03 AU. The arrival of shock on \textit{Wind} is 3.17 hr later than their arrival on \textit{STA}. We could not estimate the shock speed and the expected travel time of shock from \textit{STA} to \textit{Wind} due to the data gap in plasma measurements at the time of shock arrival.

The arrival of LE and TE on \textit{Wind} is 2.34 hr and 4.75 hr later than their arrival on \textit{STA}. Taking the measured speed of the LE and TE at \textit{STA}, it is expected that LE and TE will arrive with a delay of 2.6 hr and 2.8 hr, respectively, at \textit{Wind}. A larger delay in the arrival of the TE can be due to the expansion of the MC during its propagation from \textit{STA} to \textit{Wind} and/or a larger radial size of the cloud at \textit{Wind}. However, from in situ speed measurements, we infer that this MC has negligible expansion speed,  i.e., MC is compressed, and therefore expansion cannot explain the differences in the arrival time. It appears that even a smaller angular separation of 3.4$^\circ$ between two spacecraft has caused \textit{Wind} to observe different dimensions and regions of the MC. This implies how a single-point in situ spacecraft prevents us from understanding the global plasma properties of CMEs with inhomogeneous characteristics.

\begin{table}
    \centering
    \footnotesize
    \begin{tabular}{cccc}
    \hline
    \multicolumn{4}{c}{Arrival Time of the CME Substructures (UT)} \\
    \hline
    Substructure & STEREO-A & Wind & {$\Delta$t} (hr) \\
    \hline
   Shock & 24 Sep 17:35 & 24 Sep 20:45 & 3.17 \\
   LE & 25 Sep 06:50 & 25 Sep 09:12 & 2.34 \\
   TE & 26 Sep 06:25 & 26 Sep 11:10 & 4.75 \\
    \hline
    \multicolumn{4}{c}{Duration of the CME Substructures (hr)} \\
    \hline
   Sheath & 13.25 & 12.45 & -0.8 \\
   MC & 23.58 & 25.96 & 2.38 \\
    \hline
    \multicolumn{4}{c}{Magnitude of Total Magnetic Field During the MC (nT)} \\
    \hline
    Magnetic Field & STEREO-A & Wind & {$\Delta$B} \\
    \hline
    Maximum & 31.6 & 33.9 & 2.3 \\
    Minimum & 9.7 & 10 & 0.3 \\
    Average & 18.4 & 18.5 & 0.1 \\
    \hline
    \multicolumn{4}{c}{Magnitude of Speed During the MC (km s$^{-1}$)} \\
    \hline
    Speed & STEREO-A & Wind & {$\Delta$V} \\
    \hline
    Maximum & 476 & 493 & 17 \\
    Minimum & 399 & 378 & -21 \\
    Average & 442 & 445 & 3 \\
    \hline 
    \end{tabular}
    \caption{The first and second panels show the arrival time and duration of the CME substructures at \textit{STEREO-A}, \textit{Wind}, and the differences in measurements at both locations. The third and fourth panels show the maximum, minimum, and average magnitudes of the magnetic field and speed during the duration of the MC at \textit{STEREO-A} and \textit{Wind}, along with the differences between measurements at both locations.}
    \label{tab:tab_1}
\end{table}

The second panel of Table~\ref{tab:tab_1} shows the duration of the CME substructures, such as sheath and MC, measured on both spacecraft. We notice that the duration of the sheath/MC is less/more (-0.8 hr/2.37 hr) at \textit{Wind} in comparison to in situ observations at \textit{STA}. The third/fourth panel shows the maximum, minimum, and average magnitude of the magnetic field/speed during the duration of the MC at both spacecraft. From this, we infer that the average value of the magnetic field on \textit{STA} (18.4 nT) and \textit{Wind} (18.5 nT) are close to each other. It suggests that the MC is not expanding (i.e., compressed cloud) during its propagation of 0.03 AU ($\sim$6.4 $R_{\odot}$) distance between two spacecraft as an expanding cloud would have shown a decrease in the total magnetic field with its increasing distance \citep{CWang2005, Leitner2007, Gulisano2010, Winslow2015, EmmaDavies2021}. In the following section, we introduce a novel approach to determining the flux rope's (MC) axis and compare the estimates of the arrival of the axis center with the size center to further understand the compression of the MC.

\subsection{Proposing an Approach to Identify the Axis of the Magnetic Cloud at STEREO-A and Wind}{\label{sec:centers}}

The arrival and the orientation of the MC axis are essential parameters from the perspective of space weather. The axis center's speed and expansion govern the propagation speed of CME LE. The axis and size center are supposed to mark the center of the radial dimension of the flux rope on the in situ spacecraft. However, differences in the size and axis center arrival time can arise if the MC undergoes compression. Recently, \citet{Agarwal2024} demonstrated that larger differences in the size center from the time center (marking the center of the total duration) of the MC suggests a larger expansion speed of the MC. This implies that an MC with no expansion (i.e., compression) will have no differences in the arrival of size and time center. One can gain insights into the compression of the MC by examining the differences in the arrival time of the axis, size, and time center.

\begin{figure}
    \centering
    \includegraphics[scale= 0.41,trim={0cm 0cm 0cm 0cm},clip]{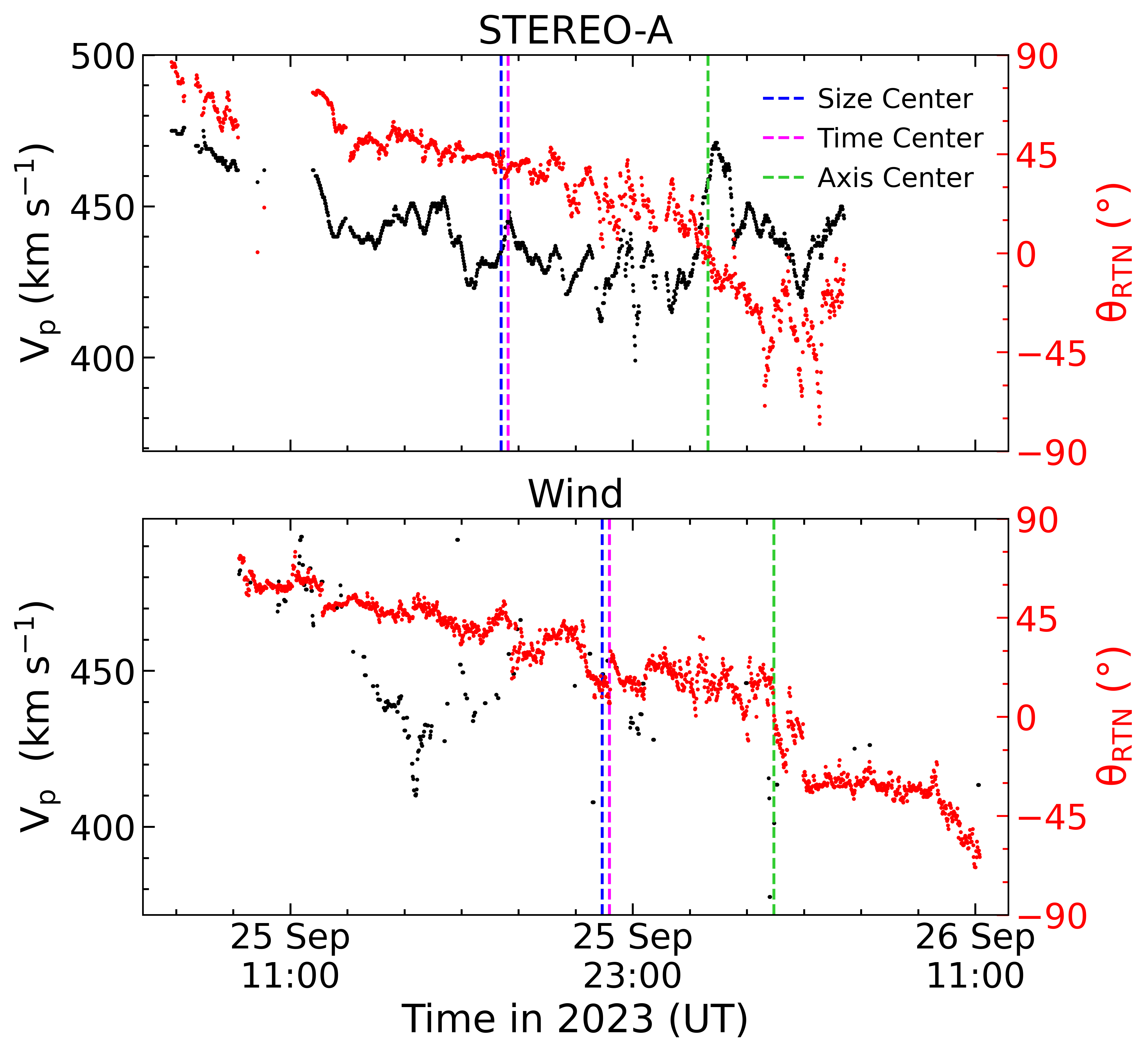}
    \caption{The top and bottom panels show the in situ measured speed profile on \textit{STA} and \textit{Wind} on the y-axis with black (left), while the y-axis with red (right) shows the variation in the latitude ($\theta$) of the total magnetic field. The blue, magenta, and green vertical dashed lines denote the MC's size center, time center, and axis center.}
    \label{fig:center}
\end{figure}

The radial size of the MC is estimated at the arrival of the LE by integrating the in situ speed with time during the duration of the MC. The estimated radial size of the MC at \textit{STA} and \textit{Wind} are 54.6 $R_{\odot}$ and 59.6 $R_{\odot}$, respectively. The arrival time of the size and time center at \textit{STA}/\textit{Wind} are 2023 September 25 at 18:23/2023 September 25 at 21:55 and 2023 September 25 at 18:37/2023 September 25 at 22:11, respectively, are listed in the top panel of Table~\ref{tab:tab_2}. The difference in the arrival of the time center from the size center at \textit{STA} and \textit{Wind} are 0.24 hr and 0.26 hr, respectively. This indicates that the MC is compressed as there is a negligible difference in the arrival time of size and time center for an MC of one-day duration \citep{Agarwal2024}. In the following, we will introduce our approach to determine the arrival of MC's axis and examine how different it is from the arrival of the size center.

\begin{figure*}
    \centering
    \includegraphics[scale= 0.37,trim={0cm 0cm 0cm 0cm},clip]{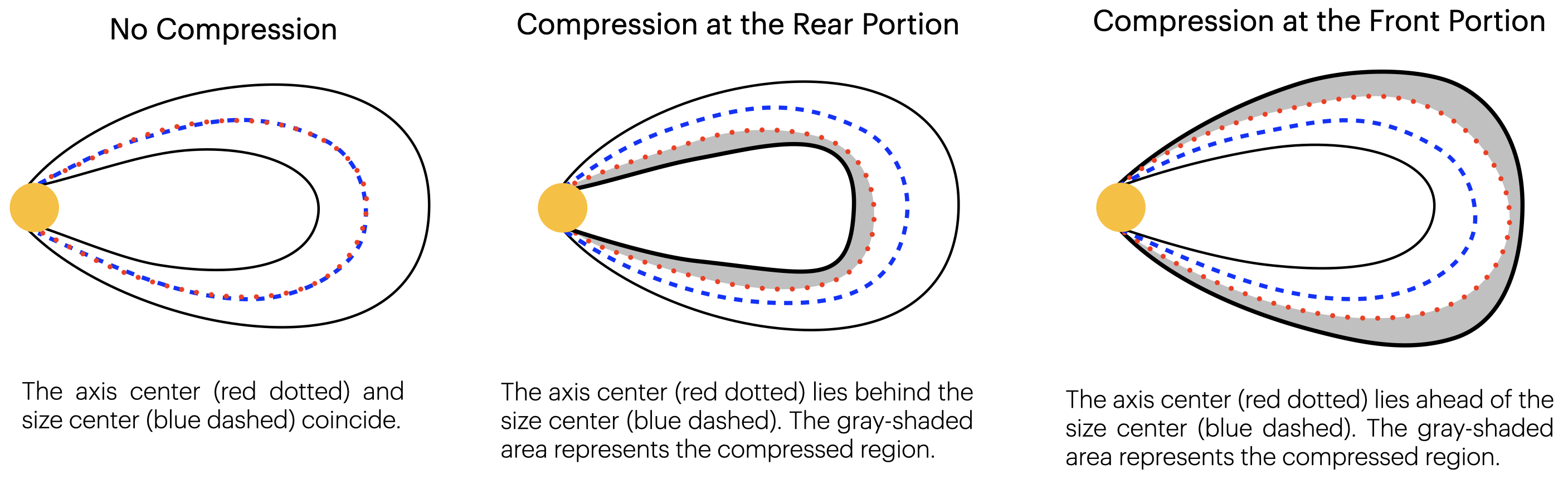}
    \caption{From the left to right, panels show the cases of MC with no compression, compression at the rear portion, and compression at the front portion. The red dotted and blue dashed lines represent the axis center and size center of the MC. The gray-shaded area represents the compressed region of the MC.}
    \label{fig:compression}
\end{figure*}

The selected MC is a low-inclined flux rope with NES orientation, which means that the normal component of the magnetic field rotates from north (N) to south (S) via east (E). In this case, the axis of the MC will be when the normal component of the magnetic field is around zero, and the flux rope is about to change its polarity from N to S. Therefore, the arrival of MC's axis (axis center) at in situ spacecraft will be at a time where the value of $\theta$ approaches nearly zero while changing its sign from '+' to '-' from MC's LE to TE. Due to fluctuations in the measurements of 1-min data, it is possible to have multiple instances of $\theta$ approaching nearly zero. Therefore, in our case, we identify the axis center across which $\theta$ shows a consistent polarity for at least 15 minutes on either side. To decide on consistency in the polarity, we require that the identified polarity be satisfied for at least 80\% intervals of the selected 15 min on either side of the MC axis.

The selected MC is a low-inclined flux rope; however, the method of determining the MC axis center would be equally valid on a highly inclined flux rope to understand its compression. For a highly-inclined flux rope, the arrival of the MC axis can be marked utilizing the longitude ($\phi$) of the magnetic field vector. For example, if the flux rope orientation is ENW, then $\phi$ can rotate from 270$^\circ$ to 90$^\circ$ and the MC's axis can be marked when the value of $\phi$ is around 180$^\circ$ in the RTN coordinate system.

The arrival times of the axis center at \textit{STA} and \textit{Wind} are 2023 September 26 at 01:38 and 2023 September 26 at 03:56, respectively, and are listed in the first panel of Table~\ref{tab:tab_2}. The axis center at \textit{Wind} arrives 2.31 hr later than at \textit{STA}. The estimated arrival of all three centers of the MC at both spacecraft are shown in Figure~\ref{fig:center}. This figure depicts the in situ speed profile in black and $\theta$ (latitude of the total magnetic field) profile in red on the left and right y-axes, respectively, for \textit{STA} and \textit{Wind} in the top and bottom panels. The size, time, and axis centers are denoted by blue, magenta, and green vertical dashed lines in both panels. From this figure and top panel of Table~\ref{tab:tab_2}, we note that the arrival of the axis center is lagging behind the size center by 7.25 hr and 6.02 hr at \textit{STA} and \textit{Wind}, respectively. In the following, we illustrate the differences in the arrival of the axis and size center with the help of a cartoon image to give a picture of the MC compression.

Figure~\ref{fig:compression} depicts the three scenarios of MC (flux rope) compression during its journey into IP medium: (i) No compression, (ii) Compression at the Rear Portion, and (iii) Compression at the Front Portion. In each scenario, the red dotted and blue dashed curved lines denote the axis and size center of the MC, respectively. The gray-shaded region represents the compression of the MC. In the first scenario, the axis and size center are the same when the MC is not compressed, and therefore, they will arrive together at the in situ spacecraft. In the cases of compression, the arrival of size and axis center will differ. As shown in the second scenario, where the MC is compressed at the rear portion, the size center will be ahead of the axis center. The third scenario is the compression of the MC from the front side; the size center will be behind the axis center. From Figure~\ref{fig:center}, it is clear that the selected MC is compressed from the rear side of both spacecraft. Compression from the rear side is also clear from the speed profile of the MC region in the fifth panel of Figure~\ref{fig:insitu_par}. The greater the compression in MC, the larger the time difference between the axis and size center. Based on the time differences between size and axis center, we note that rear side compression is more pronounced at \textit{STA} compared to \textit{Wind} even when these two spacecraft have only a small longitudinal separation of 3.4$^\circ$. In the following, we will examine the orientation of the MC's axis on both spacecraft using the MVA technique \citep{Sonnerup1967,Bothmer1998}.

\begin{table}
    \centering
    \footnotesize
    \begin{tabular}{cccccc}
    \hline
    \multicolumn{6}{c}{Arrival Time of Cloud Centers (UT)} \\
    \hline
   \multicolumn{2}{c}{Center} & {STEREO-A}  & \multicolumn{2}{c}{Wind} & {$\Delta t (hr)$}\\ 
    \hline
   \multicolumn{2}{c}{Size} & {25 Sep 18:23} & \multicolumn{2}{c}{25 Sep 21:55}& {3.54} \\
   \multicolumn{2}{c}{Time} & {25 Sep 18:37} & \multicolumn{2}{c}{25 Sep 22:11} & {3.56} \\
   \multicolumn{2}{c}{Axis} & {26 Sep 01:38} & \multicolumn{2}{c}{26 Sep 03:56} & {2.31} \\
    \hline
    \multicolumn{6}{c}{MVA Results} \\
    \hline
    \multicolumn{2}{c}{Parameters} & \multicolumn{2}{c}{STEREO-A} & \multicolumn{2}{c}{Wind} \\
    \hline
    
  \multicolumn{2}{c}{{Eigenvalues}} & \multicolumn{2}{c}{\multirow{2}{9em}{(110.22, 8.45, 3.07)}} & \multicolumn{2}{c}{\multirow{2}{9em}{(137.93, 7.27, 3.57)}}\\
  \multicolumn{2}{c}{($\lambda_1$, $\lambda_2$, $\lambda_3$)} & \multicolumn{2}{c}{}  & \multicolumn{2}{c}{} \\
  
   \multicolumn{2}{c}{$\frac{\lambda_2}{\lambda_3}$} & \multicolumn{2}{c}{2.75} & \multicolumn{2}{c}{2.04} \\
   
  \multicolumn{2}{c}{Intermediate} & \multicolumn{2}{c}{\multirow{2}{9em}{(-0.60, 0.80, 0.01)}} & \multicolumn{2}{c}{\multirow{2}{9em}{(-0.95, 0.17, 0.25)}} \\
   \multicolumn{2}{c}{Eigenvector ($e_2$)} & \multicolumn{2}{c}{} & \multicolumn{2}{c}{} \\
   
   \multicolumn{2}{c}{Orientation ($\theta$, $\phi$)} & \multicolumn{2}{c}{\multirow{2}{7em}{(-0.7$^\circ$, 307$^\circ$)}} & \multicolumn{2}{c}{\multirow{2}{7em}{(-14.5$^\circ$, 350$^\circ$)}}\\
   \multicolumn{2}{c}{of the MC Axis} & \multicolumn{2}{c}{} & \multicolumn{2}{c}{}\\
   
    \hline
    \end{tabular}
    \caption{The top panel of the table lists the arrival times of the size, time, and axis centers on \textit{STEREO-A} and \textit{Wind}, along with the differences between them. The bottom panel of the table lists the result of minimum variance analysis on \textit{STEREO-A} and \textit{Wind} during the duration of the MC.}
    \label{tab:tab_2}
\end{table}

\subsection{Orientation of the MC Axis at STEREO-A and Wind}{\label{sec:MVA}}

We determine the orientation of the axis of the MC at both spacecraft utilizing the MVA technique \citep{Sonnerup1967,Sonnerup1998,Bothmer1998,Echer2006}. The MVA method is a mathematical approach that involves determining the eigenvalues and eigenvectors of the magnetic variance matrix. The concepts and equations associated with the MVA technique are detailed in Appendix~\ref{sec:appendix} for completeness. The three eigenvalues of the magnetic variance matrix $M_{\alpha\beta}$ are $\lambda_1$, $\lambda_2$, and $\lambda_3$ (arranged in descending order) represent the actual maximum, intermediate, and minimum variances of magnetic field vector along the direction of eigenvectors $e_1$, $e_2$, and $e_3$, respectively, corresponding to each eigenvalue. The directions of three eigenvectors are mutually orthogonal to each other. The estimated directions of variances are assumed to be well-determined if $\frac{\lambda_2}{\lambda_3} \ge 2$ \citep{Siscoe1972, Lepping1980}. Therefore, the smallest eigenvalue ($\lambda_3$) is equal to the minimum variance of the magnetic field vector along the direction of the normal vector ($\hat{\textbf{n}}$ [$n_x$, $n_y$, $n_z$]), which is the direction of the eigenvector $e_3$. The direction of the MC axis is the direction of the intermediate eigenvector $e_2$.

We can compare the direction of the intermediate eigenvector or MC's axis at multiple spacecraft and analyze the differences in the estimated direction of the MC axis. The estimates of eigenvalues and the intermediate eigenvector corresponding to the MC axis from the measurements of MC at \textit{STA} and \textit{Wind} are given in the second panel of Table~\ref{tab:tab_2}. From the table, we note that the ratio of intermediate to minimum eigenvalue is 2.75 and 2.04 at \textit{STA} and \textit{Wind}, respectively. This shows that variances are well-determined, and the estimated orientation of the MC's axis at both spacecraft is reliable.

We estimate the inclination ($\theta$) and azimuthal angle ($\phi$) of the MC axis from/within the ecliptic plane at both spacecraft. For this, we use the x, y, and z components of the intermediate eigenvector ($e_2$) which can be expressed as $e_x = e_2 cos(\theta)cos(\phi)$, $e_y = e_2 cos(\theta)sin(\phi)$, and $e_z = e_2 sin(\theta)$, respectively. The value of $\phi$ ranges from 0 to 2$\pi$. We classify the value of $\phi$ into four ranges based on the sign of $e_x$ and $e_y$ simultaneously such that (i) if $e_x$ and $e_y$ are positive then the value of $\phi$ varies from 0 to $\pi/2$, (ii) if $e_x$ is negative and $e_y$ is positive then the value of $\phi$ varies from $\pi/2$ to $\pi$, (iii) if $e_x$ and $e_y$ are negative then the value of $\phi$ varies from $\pi$ to 3$\pi$/2, and (iv) if $e_x$ is positive and $e_y$ is negative then the value of $\phi$ varies from 3$\pi$/2 to 2$\pi$. The value of $\phi$, as described above, can be mathematically calculated, and it is valid for both the GSE and RTN coordinate systems. The mathematical expressions used for the calculation of $\phi$ and $\theta$ are as below:

if ${e_x>0}$ and ${e_y>0}$:
\begin{displaymath}
{\phi = \tan^{-1}\left(\frac{e_y}{e_x}\right)};~({0<\phi<\pi/2}) 
\end{displaymath}

if ${e_x<0}$ and ${e_y>0}$:
\begin{displaymath}
{\phi = \tan^{-1}\left(\frac{e_y}{e_x}\right) + 180^\circ};~({\pi/2<\phi<\pi})    
\end{displaymath}

if ${e_x<0}$ and ${e_y<0}$:   
\begin{displaymath}
{\phi = \tan^{-1}\left(\frac{e_y}{e_x}\right) + 180^\circ};~({\pi<\phi<3\pi/2})  
\end{displaymath}

if ${e_x>0}$ and ${e_y<0}$:
\begin{displaymath}
{\phi = \tan^{-1}\left(\frac{e_y}{e_x}\right) + 360^\circ};~({3\pi/2<\phi<2\pi})   
\end{displaymath}

\begin{displaymath}
\theta = tan^{-1}\left(\frac{e_z}{\sqrt{e_x^2 + e_y^2}}\right)
\end{displaymath}

The orientation ($\theta$, $\phi$) of the axis from MVA at \textit{STA} and \textit{Wind} are (0.7$^\circ$, 127$^\circ$) and (14.5$^\circ$, 170$^\circ$), respectively. The estimated value of $\theta$ at both spacecraft shows that the selected cloud of 2023 September 24-26 is a low-inclined flux rope \citep{Bothmer1998, Huttunen2005, Palmerio2018, Nieves-Chinchilla2019}. The estimated $\phi$ from MVA corresponds to the axis in the west direction at both spacecraft in the RTN system; however, the selected MC has an axis in the east direction with NES orientation as described in Section~\ref{sec:insitu}. Such ambiguity in the estimates from MVA can occur because eigenvalues of the magnetic variance matrix are always positive, as they are the actual variances in the magnetic field components \citep{Sonnerup1998,ROliveira2021} and the eigenvectors $e$ and $-e$ both are valid. To make the consistency of orientation of MC's axis estimated from MVA with in situ measurements, we added 180$^\circ$ in the estimates of $\phi$ and also reversed the sign of $\theta$ that amounts the same as reversing the sign of $e_2$ from MVA representing the MC axis. Thus, the modified and actual orientation ($\theta$, $\phi$) of the MC's axis in 3D space at \textit{STA} and \textit{Wind} are (-0.7$^\circ$, 307$^\circ$) and (-14.5$^\circ$, 350$^\circ$), respectively.

\begin{figure*}
    \centering
    \includegraphics[scale= 0.55,trim={0cm 0cm 0cm 0cm},clip]{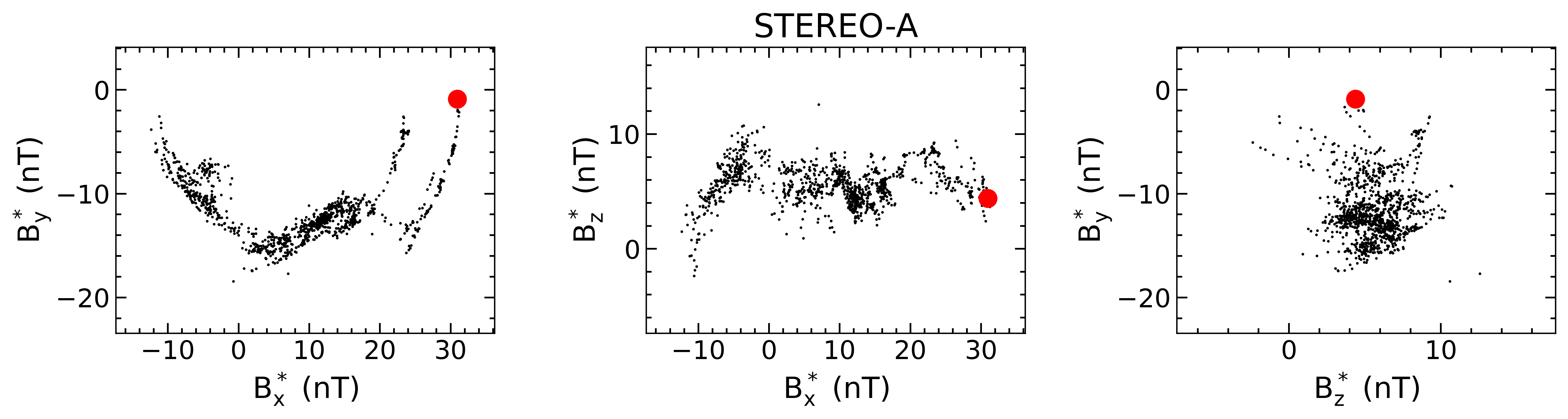}
    %\hspace*{1.5cm}
   \includegraphics[scale=0.55,trim={0cm 0cm 0cm 0cm},clip]{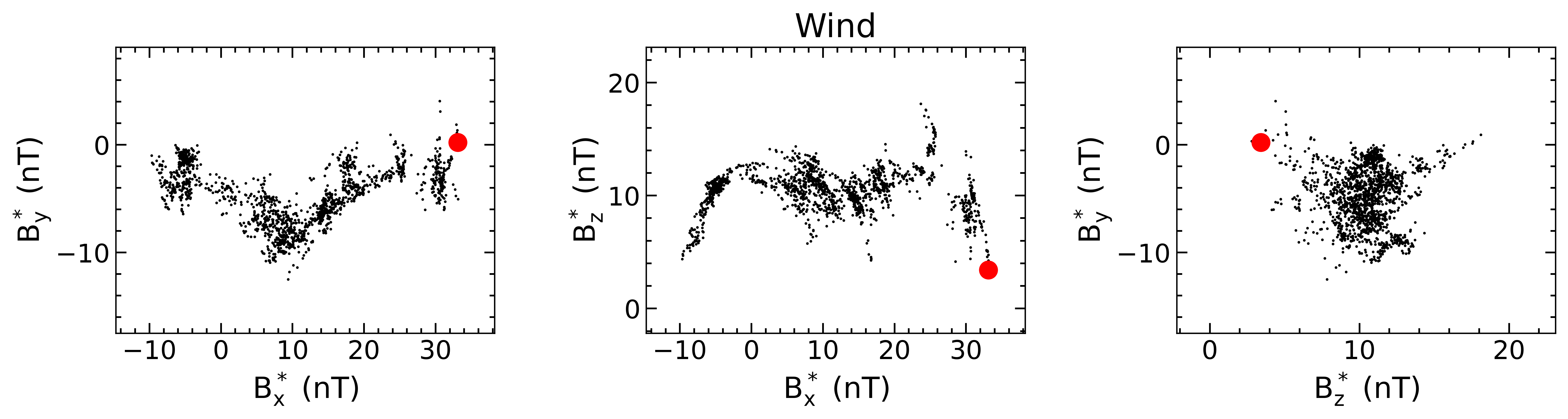}
    \caption{Hodograms representation of magnetic field vector in the direction of maximum, intermediate, and minimum eigenvectors of \textit{STA} and \textit{Wind} in the top and bottom panels, respectively. The red dot represents the starting time of the MC.}
    \label{fig:hodo}
\end{figure*}

From the estimated $\theta$, we note that the MC completely lies in the ecliptic plane for \textit{STA} with $\theta$ = -0.7$^\circ$ in compared to $\theta$ = -14.5$^\circ$ at \textit{Wind} spacecraft. The estimated value of $\phi$ shows that the orientation of the axis is not exactly aligned with the east direction ($\phi$ = 270$^\circ$) at both spacecraft. For \textit{STA}, the MC axis ($\phi$ = 307$^\circ$) is 37$^\circ$ away from the east axis, while for \textit{Wind}, the MC axis ($\phi$ = 350$^\circ$) is 80$^\circ$ away from the east axis and more towards the Sun-Earth axis. From MVA, we note the differences in the MC's axis orientation even for 3.4$^\circ$ longitudinally separated spacecraft. This suggests that there would also be differences in the orientation of vectors corresponding to maximum and minimum variances at both spacecraft. One can examine the magnetic field vector in the directions of maximum, intermediate, and minimum variances at both spacecraft using the hodogram representation, which is shown in Figure~\ref{fig:hodo}.

The top and bottom panels of the figure illustrate the hodograms representation for \textit{STA} and \textit{Wind}, respectively. The $\mathrm{B_x^*}$, $\mathrm{B_y^*}$, and $\mathrm{B_z^*}$ denotes the magnetic field vector in the direction of maximum, intermediate, and minimum eigenvectors, respectively. The red dot represents the start time of the MC. Due to the ambiguity in the sign of eigenvectors despite they satisfy a right-hand coordinate system, the hodogram representation can show different types of orientation of MC's axis. For the low inclined flux rope axis, as for the selected MC in this study, the ambiguity in eigenvectors can show four types of orientation, such as NES, SEN, SWN, and NWS, in the hodogram. For our selected MC, we have taken the sign of eigenvectors such that they become the correct choice for NES orientation as identified in Section~\ref{sec:insitu}. Such ambiguities in the hodogram and the possible solutions for them, based on the inputs of latitude and longitude of the magnetic field vector during the duration of MC, are discussed in \citet{ROliveira2021}.

The hodogram consistent with NES orientation at both spacecraft is shown in Figure~\ref{fig:hodo}. The hodogram between $\mathrm{B_y^*}$ and $\mathrm{B_x^*}$ shows a clear rotation in $\mathrm{B_x^*}$ from positive to negative as the value of $\mathrm{B_y^*}$ changes via maintaining its negative sign at both spacecraft, however; the rotation is more pronounced at \textit{STA}. The second hodogram between $\mathrm{B_z^*}$ and $\mathrm{B_x^*}$ shows the change in the value of $\mathrm{B_x^*}$ from positive to negative with the nearly constant value of $\mathrm{B_z^*}$ for \textit{STA}, while the value of $\mathrm{B_z^*}$ shows some variation at \textit{Wind}. The third hodogram between $\mathrm{B_y^*}$ and $\mathrm{B_z^*}$ shows the change in the value of $\mathrm{B_y^*}$ via maintaining its negative sign with the nearly constant value of $\mathrm{B_z^*}$ for \textit{STA}, while the value of $\mathrm{B_z^*}$ shows some variation at \textit{Wind}. This suggests that the MC axis is slightly inclined away from the ecliptic plane at \textit{Wind} but is exactly in the ecliptic plane at \textit{STA}. Two spacecraft detecting a slightly different orientation of the MC despite their much smaller separation provides insights into the local inhomogeneity in the MC either inherently present or arose during its propagation in the non-isotropic ambient medium. In the following, we will examine the magnetic field parameters on both spacecraft.

\subsection{Comparison of Magnetic Field Profiles Observed at STEREO-A and Wind}{\label{sec:mag_comp}}

\begin{figure*}
    \centering
    \includegraphics[scale= 0.82,trim={0cm 0cm 0cm 0cm},clip]{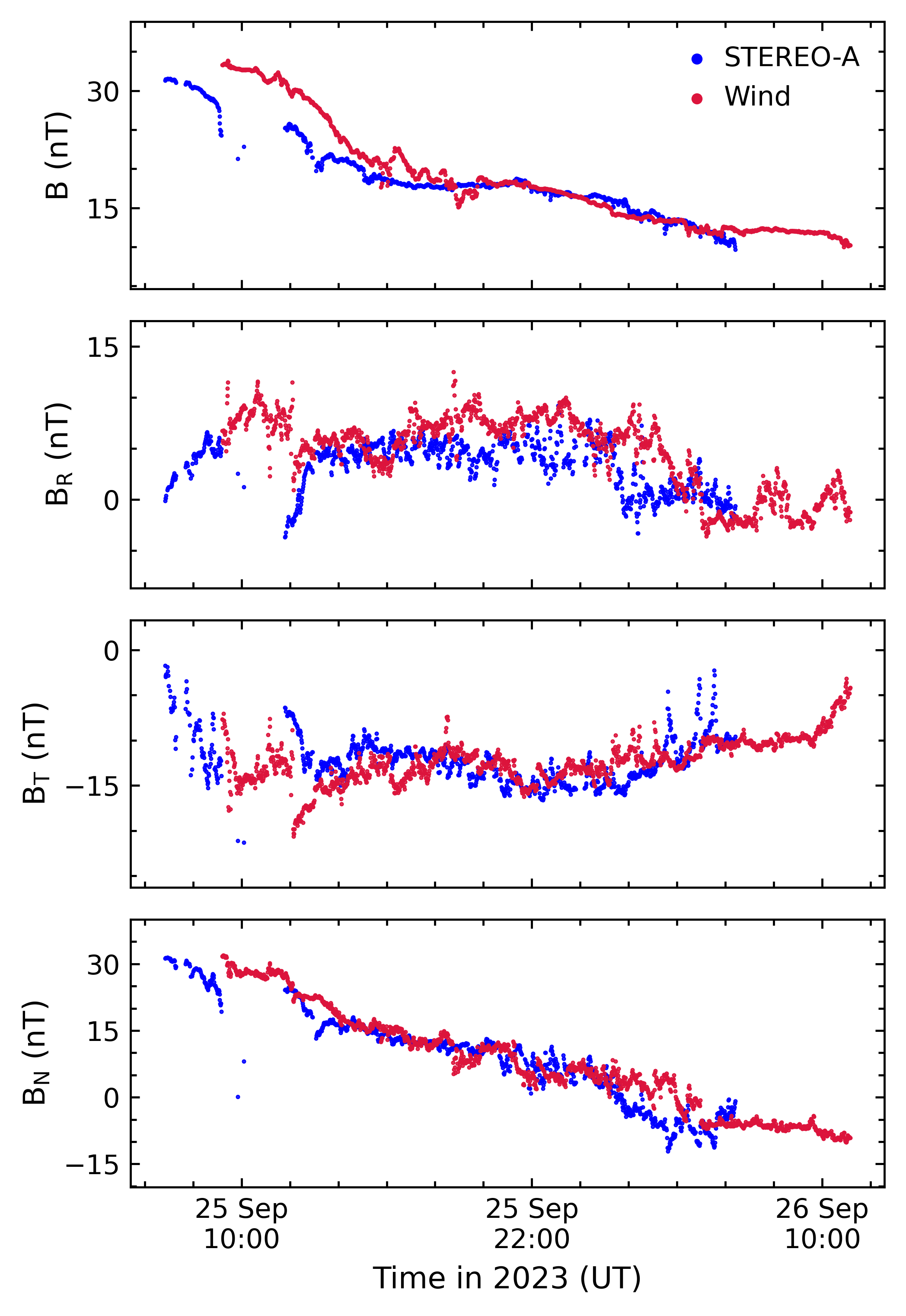}
    \includegraphics[scale= 0.82,trim={0cm 0cm 0cm 0cm},clip]{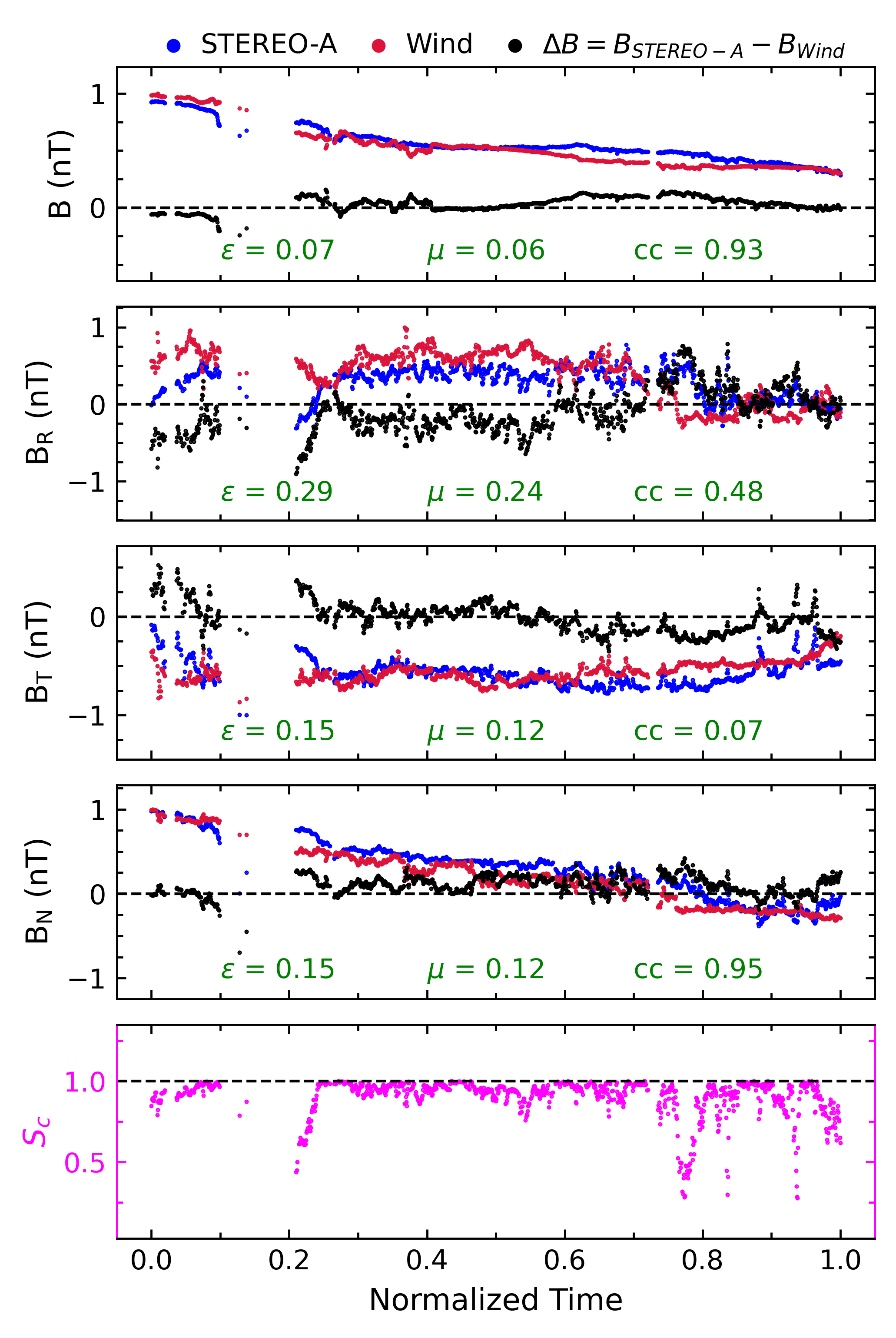}
    \caption{The left panels from top to bottom illustrate the temporal evolution of the total magnetic field and its components in the RTN coordinate system at \textit{STA} (blue) and \textit{Wind} (red). The first four panels on the right show the variations in the normalized total magnetic field and its normalized components corresponding to \textit{STA} (blue) and \textit{Wind} (red), along with their differences ($\Delta B = B_{STA} - B_{Wind}$) (black) as a function of the normalized time of the MC duration. Each panel includes the Spearman correlation coefficient (cc) between measurements at \textit{STA} and \textit{Wind} as well as root mean square error ($\epsilon$) and mean absolute error ($\mu$) for ($\Delta B$), reported in green text. The last panel of the figure shows the cosine similarity ($S_c = cos(\theta)$) between magnetic field vectors at \textit{STA} and \textit{Wind} over the normalized time of the MC duration.}
    \label{fig:mag_comp_rooterr}
\end{figure*}

We note a similar trend in the time-evolution of the total magnetic field and closely matching values of its maximum, minimum, and average at both \textit{STA} and \textit{Wind} spacecraft over the MC duration (Section~\ref{sec:insitu}). However, its RTN components at both spacecraft show noticeable differences, especially towards the trailing edge of the MC, as shown in the left panel of Figure~\ref{fig:mag_comp_rooterr}. This suggests that differences in RTN components of the field between both spacecraft may not be primarily due to its temporal evolution between both spacecraft given a radial separation of only 0.03 AU ($\sim$6.4 $R_{\odot}$) between them. Based on our earlier analysis of differing arrival times of centers and orientation of the MC at both spacecraft, we think the measured differences could be because of non-coherency in the MC structure over a small spatial scale separated by an angular extent of 3.4$^\circ$.

It is clear from the left panel of Figure~\ref{fig:mag_comp_rooterr} that both spacecraft are not sampling the LE and TE at common times and measure different durations of the MC. Therefore, we normalize the MC duration for both spacecraft to enable a one-to-one comparison between them of the temporal evolution of the magnetic field. The normalized time for the duration of the MC at both spacecraft is calculated as follows: 

$$t_{norm} = \frac{t - t_{start}}{t_{end} - t_{start}}$$

For quantifying the differences between the magnetic field parameters at both spacecraft at a common normalized time, we interpolate the measurements of \textit{Wind} (having a larger number of data points) at the normalized time of \textit{STA}. Additionally, to have an easy comparison between each magnetic field parameter, we normalize it by an absolute maximum of its measurements at both spacecraft. The right panel of Figure~\ref{fig:mag_comp_rooterr} shows the comparison of magnetic field parameters at both spacecraft with normalized time. The normalized magnetic field and its components at \textit{STA} and \textit{Wind} are shown in blue and red, respectively, in the first four panels of the figure. The difference ($\Delta B = B_{STA} - B_{Wind}$) between measurements from both spacecraft is shown in black. The horizontal black dashed line represents the zero reference to visualize the variation of $\Delta B$. We also compute the root mean square error ($\epsilon$) and mean absolute error ($\mu$) for $\Delta B$ as follows:

$${\epsilon = \sqrt{\frac{\sum_{i = 1}^N (\Delta B_i)^2}{N}}}$$
$${\mu = \frac{\sum_{i=1}^N \left|\Delta B_i\right|}{N}}$$

We note that the values of $\epsilon$ and $\mu$ are smaller for the total magnetic field compared to their values for the RTN components of the field, as reported in green text in each panel. The values of $\epsilon$ and $\mu$ are highest for the radial component of the magnetic field, while they are equal for the tangential and normal components of the magnetic field. The variation of $\Delta B$ for the total magnetic field relative to the zero reference line shows moderately non-identical values at both spacecraft for approximately one-third of the magnetic cloud (MC) during the 60\% to 90\% trailing portion of its duration. The variation of $\Delta B_{R}$ shows that $B_{R}$ at both spacecraft is significantly different for the whole duration of the MC, except for the one-tenth portion within 60\% to 70\% of the normalized MC duration. The $B_{T}$ is noted to be moderately non-identical at both spacecraft only for the first 10\% and the last 30\% portion of the MC. The $\Delta B_{N}$ and zero line are closely related during the whole duration of the MC. We also calculated the values of $\epsilon$ and $\mu$ after excluding a few visually noticeable fluctuations as outliers, especially around the data gap, and found that their values remained largely unaffected, indicating that these parameters reliably quantify the overall differences between the two profiles.

We check the Spearman (Pearson) correlation coefficient (cc) between the normalized magnetic field profiles at both spacecraft. Such an approach is also taken in \citet{Good2018}. We infer that the values of Spearman (Pearson) cc for the total magnetic field and normal component of the magnetic field are 0.93 (0.95) and 0.95 (0.94), respectively, suggesting that measurements of these parameters at both spacecraft are highly correlated with a monotonic (linear) relationship. The value of Spearman (Pearson) cc for the radial and tangential components of the magnetic field is 0.48 (0.56) and 0.07 (0.15), respectively. The low value of cc for the tangential component could also result from the smaller variations in the magnitude of $B_T$ at each spacecraft, and in such cases, the cc value may not be suitable for identifying the dissimilarity. The Spearman cc is reported in green text in each panel of the right panel of Figure~\ref{fig:mag_comp_rooterr}. Overall, the analysis suggests that the magnetic field parameters of the MC are different between \textit{STA} and \textit{Wind} measurements.

Further, we examine the directional similarity of the magnetic field vector at both spacecraft using the cosine similarity. The cosine similarity ($S_c$) is defined as a cosine of the angle between the magnetic field vector measured at \textit{STA} ($\vec{B}_{{ST-A}}$) and \textit{Wind} ($\vec{B}_{{Wind}}$) as follows:

$$S_c = cos(\theta) = \frac{\vec{B}_{{STA}} \cdot \vec{B}_{{Wind}}}{|\vec{B}_{{STA}}| |\vec{B}_{{Wind}}|}$$

The value of $S_c$ can vary from -1 to 1. A value of $S_c$ = 1, 0, and -1 implies that both vectors are exactly aligned in the same direction, orthogonal, and exactly oppositely oriented, respectively. The value of $S_c$ is shown in the bottom panel of the right-hand side Figure~\ref{fig:mag_comp_rooterr} in magenta as a function of normalized time to the MC duration. The horizontal black dashed line shows $S_c$ =1 reference for visualizing the variation above and below it. We note that the value of $S_c$ is close to 1 for most of the MC duration except for the last 20\% trailing portion of the MC. It shows that the orientation of the magnetic field vector at both spacecraft differs at the trailing portion of the MC, which is consistent with inference from $\Delta B_{R}$ and $\Delta B_{T}$. This suggests the flux rope structure is non-coherent, primarily at its trailing portion. This is also in agreement with the inferred MC compression from the rear side, as discussed in Section~\ref{sec:centers}.

\begin{figure*}
    \centering
    \includegraphics[scale= 0.92,trim={0cm 0cm 0cm 0cm},clip]{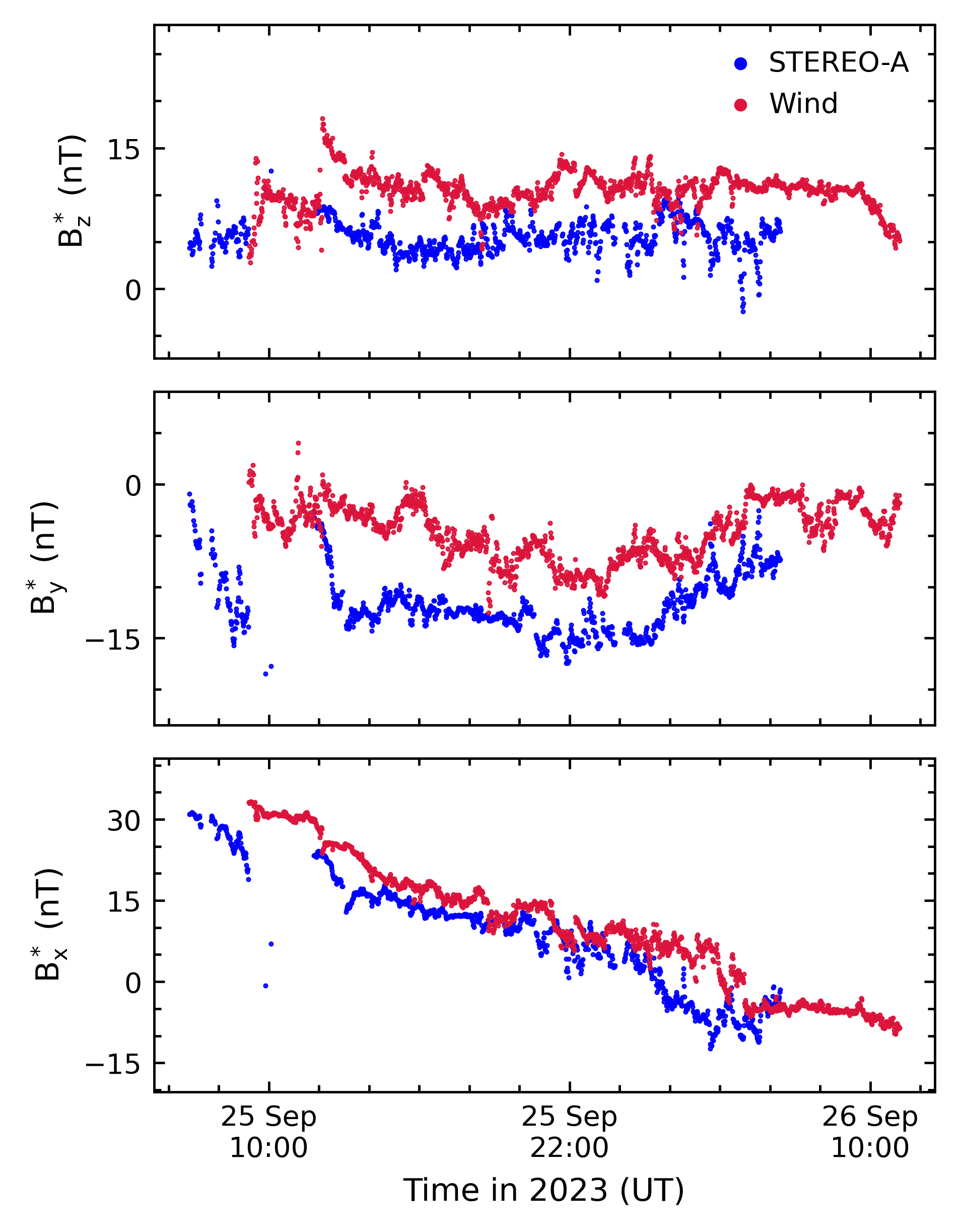}
    \includegraphics[scale= 0.92,trim={0cm 0cm 0cm 0cm},clip]{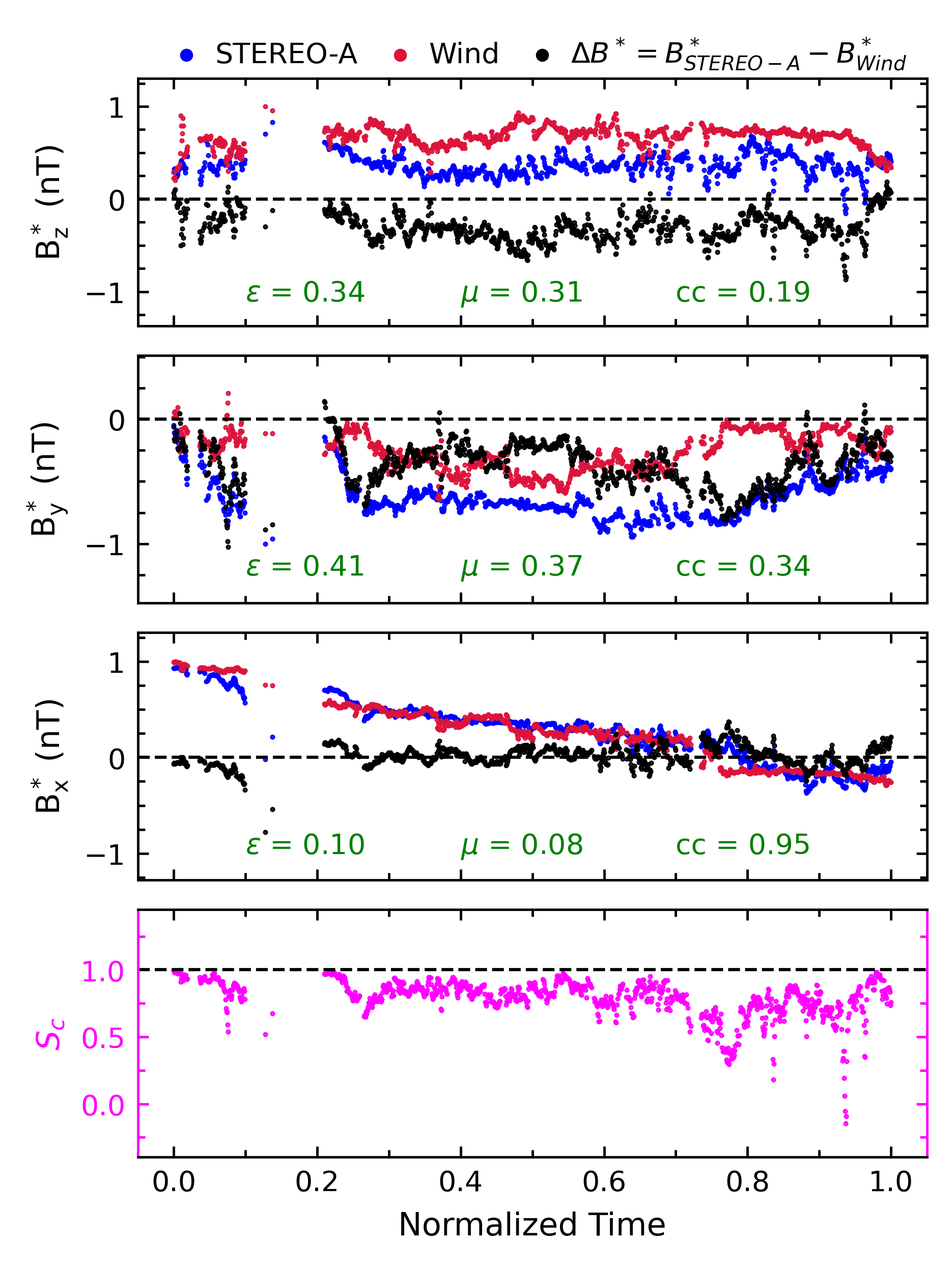}
    \caption{The left panels from top to bottom illustrate the temporal evolution of the magnetic field vector along the direction of minimum, intermediate, and maximum variance at \textit{STA} (blue) and \textit{Wind} (red). The first three panels on the right show the variations in the normalized components corresponding to \textit{STA} (blue) and \textit{Wind} (red), along with their differences ($\Delta B^* = B_{STA}^* - B_{Wind}^*$) (black) as a function of the normalized time of the MC duration. Each panel includes the Spearman correlation coefficient (cc) between measurements at \textit{STA} and \textit{Wind} as well as root mean square error ($\epsilon$) and mean absolute error ($\mu$) for ($\Delta B^*$), reported in green text. The last panel of the figure shows the cosine similarity ($S_c = cos(\theta)$) between magnetic field vectors, based on its components, along the direction of minimum, intermediate, and maximum variance at \textit{STA} and \textit{Wind} over the normalized time of the MC duration.}
    \label{fig:mag_comp_rooterr_MVA}
\end{figure*}

We also attempt to investigate the differences in the magnetic field components along the directions of minimum ($B_z^*$), intermediate ($B_y^*$), and maximum ($B_x^*$) variance (from MVA) at \textit{STA} and \textit{Wind}, as shown in the left panel of Figure~\ref{fig:mag_comp_rooterr_MVA}. In the right panel of the figure, the first three panels show the normalized field parameters along the direction of different variance axes for \textit{STA} (blue) and \textit{Wind} (red). The difference ($\Delta B^* = B_{STA}^* - B_{Wind}^*$) in these parameters are shown in black. The normalized components of the field and time are estimated using a similar approach as described above for the measured magnetic field components in the RTN coordinate system. The estimated values of $\epsilon$ and $\mu$ for the difference in the components are reported in green text in each panel. From this, we note that these values are smallest for $\Delta B_{x}^*$ in comparison to $\Delta B_{y}^*$ and $\Delta B_{z}^*$. However, the values of $\epsilon$ and $\mu$ are largest for $\Delta B_{y}^*$. This suggests that the direction of the MC axis at both spacecraft is different, even when the angular separation between spacecraft is 3.4$^\circ$, which is expected from Section~\ref{sec:MVA}. This also shows that the direction of minimum variance within the MC is significantly different for both spacecraft.

The horizontal black dashed line in the figure represents the zero reference line for $\Delta B^*$. From this, we infer that the $B_z^*$ and $B_y*$ are different at both spacecraft during the complete duration of the MC. However, the $B_x^*$ at both spacecraft is almost similar during the MC duration. The values of Spearman (Pearson) cc for $B_z^*$, $B_y^*$, and $B_x^*$ are 0.19 (0.13), 0.34 (0.34), and 0.95 (0.95). The value of cc is highest for the $B_x^*$ in agreement with the smallest value of $\epsilon$ and $\mu$. The low value of cc for the $B_z^*$ and $B_y^*$ could be because of the larger values of $\epsilon$ and $\mu$ as well as relatively constant magnitudes of $B_z^*$ and $B_y^*$ at both spacecraft. We also estimate the $S_c$ (for directional similarity) between magnetic field vectors at both spacecraft comprising $B_z^*$, $B_y^*$, and $B_x^*$. The horizontal black dashed line represents the value of $S_c$ = 1. We infer that the direction of the total magnetic field vector, based on its components along the direction of variances, at both spacecraft is non-identical during the last 40\% duration of MC. This can happen due to the compression of the MC from the rear side, as discussed in Section~\ref{sec:centers}.

\section{RESULTS AND DISCUSSION}{\label{sec:result}}

The present study focuses on analyzing the observations of a selected CME at two closely spaced in situ spacecraft near 1 AU and examining the differences between the measured characteristics of the CME at both locations. The selected CME was observed at \textit{STA} and \textit{Wind} spacecraft during 2023 September 24-26, with minimal radial and longitudinal separations between spacecraft of 0.03 AU and 3.4$^\circ$, respectively. Such unique locations of both spacecraft place them in a rare conjunction and enable the mesoscale observations of the CME. We estimate the arrival times of CME substructures (shock, LE, and TE), MC axis, and MC orientation at \textit{STA} and \textit{Wind} and compare the estimates for both locations. We also compare the trends of variations in the magnetic field parameters within the MC duration for both locations. Such a comparison of multipoint measurements at a small scale provides insights into inhomogeneity within CME, which could be a consequence of its kinematic deformation or its non-coherent structure during its propagation in the non-uniform solar wind \citep{Owens2017,Al-haddad2025}.

We determine the MC's axis (i.e., axis center) by employing a relatively unexplored method that utilizes the variations in the latitude/longitude of the total magnetic field over the MC duration. The arrival time estimates of the axis center and size center of the MC at any spacecraft help us to understand its compression at the front/rear portion of the MC. Moreover, comparing these centers on both longitudinally separated spacecraft can shed light on the compression of the MC along its angular extent. The asynchrony between the axis and size centers of the MC at both spacecraft suggests the compression of the MC (Figure~\ref{fig:center}). The axis center is lagging behind the size center of the MC by 7.25 and 6.02 hr at \textit{STA} and \textit{Wind}, respectively. If there is a larger time difference between the arrival of the axis and the size center, it would suggest a more substantial compression of the CME, resulting in its deformation. This suggests that the MC at both spacecraft is compressed from the rear side, which is stronger at \textit{STA} compared to \textit{Wind}. The compression of CMEs/MCs has also been previously noted because of their interaction with following faster solar wind, CMEs, and shocks \citep{Mishra2014,Temmer2014,Mishra2015a,Heinemann2019}. The compression of the selected MC is also evident as the time center and size center are synchronous at both spacecraft (Figure~\ref{fig:center}). The compression of the MC inferred from the asynchrony between the axis and size centers can be illustrated in Figure~\ref{fig:compression}. The study of \citet{Winslow2021} has reported the compression in an MC observed at \textit{PSP} and \textit{STA} that were separated radially and longitudinally by 0.19 AU and 8$^\circ$, respectively. Their study did not find the compression of the MC along its angular extent. However, we find evidence of non-isotropic compression for the selected MC and suggest such a possibility even at the mesoscale.

The multipoint in situ measurements of the selected CME/MC show that the arrival of its substructures (shock, LE, and TE) at \textit{Wind} is later than their arrival at \textit{STA} due to the small radial separation of 0.03 AU ($\sim$6.4 $R_{\odot}$) between both spacecraft. The arrival of the TE at \textit{Wind} is significantly later than the arrival of other substructures. This implies a larger radial size of the MC at \textit{Wind} (59.6 $R_\odot$) in comparison to \textit{STA} (54.6 $R_\odot$), possibly due to a relatively weaker compression or lesser deformation at \textit{Wind}. Our finding suggests that even a smaller angular separation of 3.4$^\circ$ between two spacecraft can cause them to detect radial sizes differing by 10\%. Our finding is important as both spacecraft are almost in the same plane, and they are likely to have the same crossing distance from the MC axis. The differences in the detected radial sizes of the CMEs/MCs at multiple spacecraft have been noted in earlier studies, particularly for spacecraft having larger radial/longitudinal separations between them \citep{Crooker1996,Leitner2007,Lugaz2020}.

We further employ the MVA on magnetic field parameters and determine the orientation of MC's axis at both spacecraft. The orientation ($\theta$, $\phi$) of the axis at \textit{STA} and \textit{Wind}, consistent with NES orientation, are (-0.7$^\circ$, 307$^\circ$) and (-14.5$^\circ$, 350$^\circ$), respectively. The small angular separation of even 3.4$^\circ$ between two in situ spacecraft could show such a difference in MC's axis orientation. The orientation of MC is slightly out-of-ecliptic at \textit{Wind}, and its effect reduces the sampled radial size \citep{Crooker1996,Bothmer1998}, which contrasts our finding. Therefore, we think that the effect of non-isotropic compression is larger than the orientation of the MC axis in governing the measured radial sizes at both spacecraft. From the hodogram representation (Figure~\ref{fig:hodo}) consistent with NES orientation, we note a clear rotation in $\mathrm{B_x^*}$ for hodogram between $\mathrm{B_y^*}$ and $\mathrm{B_x^*}$ at both spacecraft; however, the rotation is more evident for \textit{STA}. The hodogram between $\mathrm{B_z^*}$ and $\mathrm{B_x^*}$ depicts a minimal rotation in $\mathrm{B_x^*}$ only for \textit{Wind} not for \textit{STA} which is expected. Similar to our study, \citet{Liu2008} also note the discernible differences in the orientation of the 2007 May MC observed by the twin \textit{STEREO} and \textit{Wind} having a small angular separation between them.

The comparison between magnetic field parameters at both spacecraft shows the noticeable differences in radial and tangential components of the magnetic field, especially during the trailing portion of the MC. Also, the cosine similarity analysis suggests that the direction of the magnetic field vector at both spacecraft is different towards the trailing part of the MC. This could be possible due to the compression of the MC from the rear side. The total magnetic field and its normal component at both spacecraft exhibit a strong correlation. Additionally, the magnetic field vector along the direction of the intermediate variance axis at both spacecraft shows significant differences in comparison to the magnetic field vector along the direction of minimum and maximum variance, consistent with the axis orientation from the MVA analysis. The cosine similarity analysis between the magnetic field vector along the direction of variance axes at both spacecraft reveals the vectors are in different directions, significantly in the trailing part of the MC. From this analysis, we identify non-coherency in the magnetic field structure in the trailing part of the MC, it could be possible because of compression in the rear side or can be inherent. However, we note that our analysis can have biases due to the inherent limitations of 1D in situ measurements, which have difficulty capturing the full complexity of a 3D structure of an MC. Also, the findings from the MVA method have uncertainties due to changes in the identified boundaries and duration of the MC \citep{ROliveira2021}.

Our study established that even for a small angular separation of 3.4$^\circ$ between the multipoint in situ spacecraft, the plasma and magnetic field parameters of an MC are anisotropic or inhomogeneous at mesoscale along its angular extent. This finding could be due to the influence of the non-uniform ambient medium on the MC or its inherent non-coherent flux rope structure in the MC itself. We emphasize that plasma measurements of the MC taken by a single point in situ spacecraft could not represent the global properties of CMEs. Similar to our study, earlier studies also reported differences in the magnetic field properties of the MC along its small angular extent \citep{Kilpua2011,Lugaz2018}. Also, the study of \citet{Regnault2024} found notable changes in both the magnetic field and the speed of an MC observed by \textit{SolO} and \textit{Wind}, which were separated radially by 0.13 AU and longitudinally by 2.2$^\circ$. Our study additionally notes a non-isotropic compression along the angular extent of the MC, which is not present in the case study of \citet{Regnault2024}. Our findings offer an alternative perspective to that of \citet{Winslow2021}, which reported a uniform distortion of an MC over a comparatively larger angular extent of around 8$^\circ$. Our study contributes a case study to the existing literature, supporting the idea that MC observations observed by a single in situ spacecraft reflect local characteristics rather than global ones \citep{Mostl2012,Mishra2021,Lugaz2022}.

The study of \citet{EmmaDavies2020} found that the MC properties measured from near-Earth spacecraft (\textit{Wind}, \textit{ACE}, \textit{THEMIS B}, and \textit{THEMIS C} having mutual separations smaller than 0.01 AU and 0.2$^\circ$) have no considerable changes in the magnetic field parameters of MC along its small angular extent. Therefore, in the context of our study, it would be important to understand the scales over which the MC characteristics could be assumed to be the same. The observed differences in the orientation of MC at various locations suggest that longitudinal separation, rather than radial separation between the spacecraft, predominantly influences the magnetic field properties. Our study highlights that a few MCs could exhibit inhomogeneity in their properties along a small angular extent.

Our study shows the mesoscale differences of the CME observed on 2023 September 24-26 at \textit{STA} and \textit{Wind} near 1 AU distance from the Sun. This study shows the necessity of more multipoint in situ spacecraft studies along the angular extent of the CME to further understand the physical processes responsible for inhomogeneity at different scales within CME. Future studies making statistics of such CMEs observed by multiple spacecraft separated by much smaller distances and longitudes will help understand if most CMEs have appreciable inhomogeneity at mesoscales because of the flux rope's non-coherent structure. Also, the role of interactions between CMEs and surrounding non-uniform ambient medium needs to be examined to get insights into the growth of an existing inhomogeneity in CMEs at different scales. Therefore, a comprehensive understanding of the global characteristics of MCs requires coordinated multi-spacecraft observations of the same MC across its angular extent from regions near the Sun through interplanetary space.

\section*{Acknowledgement}
We thank the STEREO (IMPACT and PLASTIC) and Wind (MFI and SWE) spacecraft team members for making their observations publicly available. We also thank the anonymous referee for his/her careful review that improved the manuscript.

\vspace{5mm}
\facilities{STEREO-A (IMPACT and PLASTIC) and Wind (MFI and SWE)}

\appendix

\section{Minimum Variance Analysis}\label{sec:appendix}

The MVA technique is used to determine the direction in space along which the magnetic field vector over the complete duration of the MC  has a minimum variance. This implies that the technique determines the normal vector ($\hat{\textbf{n}}$ [$n_x$, $n_y$, $n_z$]) such that the in situ measured magnetic field vector ($\textbf{B}^{(i)}$) projected along the normal direction ($\textbf{B}^{(i)}\cdot \hat{\textbf{n}}$) will have minimum variance, where \textbf{B} represents the measured components of the magnetic field vector ($B_x$, $B_y$, $B_z$) in the cartesian coordinate system (e.g., GSE or RTN) at times $i = 1,2,......,N$ during the passage of the MC on the spacecraft. The minimization of variance estimates the normal vector for the magnetic field vectors along the normal direction ($\textbf{B}^{(i)}\cdot \hat{\textbf{n}}$) under the constraint $|\hat{\textbf{n}}|^2 = 1$. Thus, the variance can be written as follows:

\begin{equation}\label{equ:1}
\sigma^2 = \frac{1}{N} \sum_{i = 1}^{N} |(\textbf{B}^{(i)} - \langle \textbf{B} \rangle) \cdot \hat{\textbf{n}}|^2
\end{equation}

where $\langle \textbf{B} \rangle$ represents the average of $\textbf{B}^{(i)}$ over the duration of the MC, i.e., 
$$\langle \textbf{B} \rangle = \frac{1}{N} \sum_{i = 1}^{N} \textbf{B}^{(i)}$$

The minimization of variance under the constraint of the normal vector ($|\hat{\textbf{n}}|^2 -1$ = 0) has been estimated using a Lagrange multiplier $\lambda$. This reduces to estimating the solution of three homogeneous linear equations~\ref{equ:2}, \ref{equ:3} and \ref{equ:4} as given below:

\begin{equation}\label{equ:2}
    \frac{\partial}{\partial n_x}\left[\sigma^2 - \lambda(|\hat{\textbf{n}}|^2 - 1)\right] = 0
\end{equation}

\begin{equation}\label{equ:3}
    \frac{\partial}{\partial n_y}\left[\sigma^2 - \lambda(|\hat{\textbf{n}}|^2 - 1)\right] = 0
\end{equation}

\begin{equation}\label{equ:4}
    \frac{\partial}{\partial n_z}\left[\sigma^2 - \lambda(|\hat{\textbf{n}}|^2 - 1)\right] = 0
\end{equation}

We further expand equation~\ref{equ:1} for completeness in the literature as follows,

\begin{align*}
\sigma^2 =  \frac{1}{N} \sum_{i = 1}^{N} [({B_x}^{(i)} - \langle {B_x} \rangle)^2 {{n_x}}^2 + ({B_y}^{(i)} - \langle {B_y} \rangle)^2{{n_y}}^2\\
 + ({B_z}^{(i)} - \langle {B_z} \rangle)^2 {{n_z}}^2\\ 
 +  2({B_x}^{(i)} - \langle {B_x} \rangle)({B_y}^{(i)} - \langle {B_y} \rangle) {{n_x}}{{n_y}}\\ 
 + 2({B_x}^{(i)} - \langle {B_x} \rangle)({B_z}^{(i)} - \langle {B_z} \rangle) {{n_x}}{{n_z}}\\ 
 + 2({B_y}^{(i)} - \langle {B_y} \rangle)({B_z}^{(i)} - \langle {B_z} \rangle) {{n_y}}{{n_z}}]\\
\end{align*}

\begin{align}\label{equ:5}
\sigma^2 =  [(\langle {B_x B_x} \rangle - \langle {B_x} \rangle \langle {B_x} \rangle) {{n_x}}^2 \nonumber\\
 + (\langle {B_y B_y} \rangle - \langle {B_y} \rangle \langle {B_y} \rangle){{n_y}}^2 \nonumber\\ 
 + (\langle {B_z B_z} \rangle - \langle {B_z} \rangle \langle {B_z} \rangle) {{n_z}}^2 \nonumber\\ 
 + 2(\langle {B_x B_y} \rangle - \langle {B_x} \rangle \langle {B_y} \rangle) {{n_x}}{{n_y}} \nonumber\\ 
 + 2(\langle {B_x B_z} \rangle - \langle {B_x} \rangle \langle {B_z} \rangle) {{n_x}}{{n_z}} \nonumber\\ 
 + 2(\langle {B_y B_z} \rangle - \langle {B_y} \rangle \langle {B_z} \rangle) {{n_y}}{{n_z}}]
\end{align}

Using equation~\ref{equ:5}, equation~\ref{equ:2}, \ref{equ:3} and \ref{equ:4} can also be written as follows:

\begin{align}\label{equ:6}
(\langle {B_xB_x} \rangle - \langle {B_x} \rangle  \langle {B_x} \rangle){n_x} + (\langle {B_xB_y} \rangle - \langle {B_x} \rangle  \langle {B_y} \rangle){n_y}\nonumber\\+ (\langle {B_x B_z} \rangle - \langle {B_x} \rangle \langle {B_z} \rangle){n_z} = \lambda n_x
\end{align}

\begin{align}\label{equ:7}
(\langle {B_xB_y} \rangle - \langle {B_x} \rangle  \langle {B_y} \rangle){n_x} + (\langle {B_yB_y} \rangle - \langle {B_y} \rangle  \langle {B_y} \rangle){n_y}\nonumber\\+ (\langle {B_y B_z} \rangle - \langle {B_y} \rangle \langle {B_z} \rangle){n_z} = \lambda n_y
\end{align}

\begin{align}\label{equ:8}
(\langle {B_xB_z} \rangle - \langle {B_x} \rangle  \langle {B_z} \rangle){n_x} + (\langle {B_yB_z} \rangle - \langle {B_y} \rangle  \langle {B_z} \rangle){n_y}\nonumber\\+ (\langle {B_z B_z} \rangle - \langle {B_z} \rangle \langle {B_z} \rangle){n_z} = \lambda n_z
\end{align}

The above equations \ref{equ:6}, \ref{equ:7}, and \ref{equ:8} can be represented in a matrix form as shown below:

\begin{equation}\label{equ:9}
\sum_{\beta = 1}^3 M_{\alpha \beta} n_{\beta} = \lambda n_{\alpha}
\end{equation}

where $M_{\alpha\beta} = \langle {B_\alpha B_\beta} \rangle - \langle {B_\alpha} \rangle \langle {B_\beta} \rangle$ and $\alpha, \beta$  = 1, 2, 3 which represent the x, y, z components of the cartesian coordinate system.

% \bibliography{anjali_ref_merge}{}
% \bibliographystyle{aasjournal}

\end{document}